\documentclass[aps,preprint,showkeys,superscriptaddress,nofootinbib,floatfix]{revtex4-1}
\usepackage{amsmath}
\usepackage[pdftex]{graphicx}  
\usepackage{slashed}
\usepackage{newtxtext,newtxmath}
\usepackage{xcolor}
\newcommand{\1}{\mbox{1}\hspace{-0.25em}\mbox{l}}
\makeatletter

\@addtoreset{equation}{section}

\begin{document}

\title{\bf Four quark operators for kaon bag parameter with gradient flow}

\preprint{UTHEP-750, UTCCS-P-133, KYUSHU-HET-213}

\author{Asobu Suzuki}
\email{suzuki@het.ph.tsukuba.ac.jp}
\affiliation{Institute of Physics, University of Tsukuba, Tsukuba, Ibaraki 305-8571, Japan}
\author{Yusuke Taniguchi}
\email{tanigchi@het.ph.tsukuba.ac.jp}
\affiliation{Center for Computational Sciences, University of Tsukuba, Tsukuba, Ibaraki 305-8577, Japan}
\author{Hiroshi Suzuki}
\email{hsuzuki@phys.kyushu-u.ac.jp}
\affiliation{Department of Physics, Kyushu University, 744 Motooka, Nishi-ku, Fukuoka 819-0395, Japan}
\author{Kazuyuki Kanaya}
\email{kanaya@ccs.tsukuba.ac.jp} 
\affiliation{Tomonaga Center for the History of the Universe, University of Tsukuba, Tsukuba, Ibaraki 305-8571, Japan}

\date{\today}

\begin{abstract}
To study the \textit{CP}-violation using the~$K_0-\overline{K}_0$ oscillation, we need the kaon bag parameter which represents QCD corrections in the leading Feynman diagrams. 
The lattice QCD provides us with the only way to evaluate the kaon bag parameter directly from the first principles of QCD.
However, a calculation of relevant four quark operators with theoretically sound Wilson-type lattice quarks had to carry a numerically big burden of extra renormalizations and resolution of extra mixings due to the explicit chiral violation.
Recently, the Small Flow-\textit{t}ime eXpansion (\textbf{SF\textit{t}X}) method was proposed as a general method based on the gradient flow  to correctly calculate any renormalized observables on the lattice, irrespective of the explicit violations of related symmetries on the lattice.
To apply the SF\textit{t}X method, we need matching coefficients, which relate finite operators at small flow times in the gradient flow scheme to renormalized observables in conventional renormalization schemes.
In this paper, we calculate the matching coefficients for four quark operators and quark bilinear operators, relevant to the kaon bag parameter.
\end{abstract}

\maketitle

\section{Introduction}

In the study of the \textit{CP}-violation, the~$K_0-\overline{K}_0$ oscillation plays an important role. 
Here, to extract Cabibbo-Kobayashi-Maskawa matrix elements in the leading Feynman diagrams for the~$K_0-\overline{K}_0$ oscillation, we need to know the kaon bag parameter which represents QCD corrections in these diagrams.
The lattice QCD provides us with the only way to evaluate the nonperturbative value of the kaon bag parameter directly from the first principles of QCD~\cite{history}.
However, a calculation of relevant four quark operators with theoretically sound Wilson-type lattice quarks had to carry a numerically big burden of extra renormalizations  and resolution of extra mixings, required mainly due to the explicit violation of the chiral symmetry by the Wilson quarks at nonzero lattice spacings~\cite{BK_01,BK_02,BK_03,BK_04,BK_05,BK_06,BK_07,BK_08,BK_09,BK_10}.

Recently, a series of new methods based on the gradient flow introduced various advances in lattice QCD~\cite{GF_00,GF_03,GF_04,GF_01,GF_02,GF_05,GF_ap01,GF_ap03,GF_10,GF_11,GF_12}. 
Among them, we adopt the \textbf{S}mall  \textbf{F}low-\textbf{\textit{t}}ime e\textbf{X}pansion (\textbf{SF\textit{t}X}) method, which is a general method to correctly calculate any renormalized observables on the lattice~\cite{GF_ap01,GF_ap03,Tec_01,Tec_02,Tec_04}.  
The gradient flow is a modification of bare fields according to flow equations driven by the gradient of an action. 
It is shown that the operators constructed by flowed fields (``flowed operators'') are free from UV divergence and also from short-distance singularities at nonzero flow time~$t>0$~\cite{GF_01}.
The basic idea of the SF\textit{t}X method is as follows:
Because of the strict finiteness of flowed operators, we can safely evaluate their nonperturbative values by (i) constructing their lattice operators directly from their continuum expressions, (ii) evaluating their values on the lattice, and (iii) taking the continuum limit. 
The finiteness of the target operators leads us automatically to their correct values by just taking the continuum extrapolation---we do not need to introduce any additional corrections due to the lattice artifacts, even if the lattice model at finite lattice spacings violates some symmetries relevant to the original derivation of the operators. 

The method has been applied to calculate the energy-momentum tensor, which is the generator of the continuous Poincar\'{e} transformation and thus is not straightforward to evaluate on discrete lattices.
From test studies around the deconfinement transition temperature in quenched QCD~\cite{Asakawa:2013laa,FlowQCD1,Iritani2019} and in $2+1$ flavor QCD with improved Wilson quarks~\cite{GF_ap08,GF_ap09,Lat2017-kanaya,EMT-2loop1,EMT-2loop2}, it was shown that the results of 
the energy-momentum tensor by the SF\textit{t}X method correctly reproduce previous results of the equation of state estimated by the conventional integral methods.

Because the method is applicable also to observables related to the chiral symmetry, we may apply the method to cope with the difficulties of Wilson-type quarks associated with their explicit chiral violation.
Theoretical basis to study fermion bilinear operators in the SF\textit{t}X method is given in~\cite{Tec_04}. 
The method was applied to compute the disconnected chiral susceptibility in $2+1$ flavor QCD with improved Wilson quarks~\cite{GF_ap08}. 
It was shown that the chiral condensates bend sharply and the disconnected chiral susceptibilities show peak at the  pseudocritical temperature.
The method was further applied to compute topological susceptibilities using the gluonic and fermionic definitions~\cite{GF_ap09}.
In the continuum, the two definitions should lead to the same results thanks to a chiral Ward-Takahashi identity, but, they are largely discrepant with the conventional lattice method at nonzero lattice spacings.
With the SF\textit{t}X method, the two definitions are shown to agree well with each other even at a finite lattice spacing~\cite{GF_ap09}.
These suggest that the SF\textit{t}X method is powerful in calculating correctly renormalized observables.

In this paper, we extend the SF\textit{t}X method to the study of four fermi operators. 
As the first step of the study, we concentrate on the issue of the kaon bag parameter,
\begin{eqnarray}
B_K&=&\frac{\langle \overline{K^0} | \, O^{\Delta{S}=2} \, | K^0 \rangle }{\frac{8}{3}\left| \langle 0 | \, \overline{s}\, \gamma_{\mu}\gamma_{5}\, d \, | K^0 \rangle \right|^2} ,
\label{eq:BK}
\end{eqnarray}
where, with $\gamma^{L}_{\mu} := \gamma_{\mu} \,(1-\gamma_5) $,
\begin{eqnarray}
O^{\Delta{S}=2} &=& (\overline{s}\,\gamma^{L}_{\mu}\,d)(\overline{s}\,\gamma^{L}_{\mu}\,d)
\label{eq:DeltaS2}
\end{eqnarray}
is the~$\Delta{S}=2$ four quark operator. 
In a conventional lattice calculation with Wilson-type quarks, due to the violation of the chiral symmetry, this four quark operator is contaminated by other operators which have the same parity and different chirality:
$
O^{\Delta{S}=2}_\text{Ren.} = ZO^{\Delta{S}=2}+\sum_{i}Z_{i}O_{i}.
$
Precise evaluation of the renormalization and mixing coefficients is computationally demanding~\cite{BK_11}.
In a real scalar field theory, the gradient flow was shown to avoid the issue of operator mixing~\cite{sOPE}. See also recent studies~\cite{GF_ap10,GF_ap11,GF_ap12,GF_ap13}.
We thus expect that the SF\textit{t}X method will drastically simplify the calculation of four quark operators in QCD%
\footnote{See Refs.~\cite{Twisted_mass01,Twisted_mass02} for a different approach using twisted mass Wilson-type quarks.
}. 

The SF\textit{t}X method~\cite{GF_ap01,GF_ap03,Tec_01,Tec_02,Tec_04} is based on the expansion of flowed operators at small~$t$ in terms of renormalized operators at $t=0$ in a conventional renormalization scheme, say the $\overline{\textrm{MS}}$ scheme~\cite{GF_01}.
The coefficients relating both operators are called the matching coefficients.
Because the renormalization scale for flowed operators can be taken to be proportional to $1/\sqrt{t}$, in asymptotically free theories such as QCD, we can calculate the matching coefficients at small~$t$ by perturbation theory.
In this paper, we perform a one-loop calculation of the matching coefficient for the $\Delta{S}=2$ four quark operator~(\ref{eq:DeltaS2}).

This paper is organized as follows:
In Sec.~\ref{sec:formulation}, we introduce the gradient flow and the dimensional reduction scheme we adopt. 
As shown in Eq.~(\ref{eq:BK}), the kaon bag parameter consists of a four quark operator and a quark bilinear operator in the denominator.
We study the matching coefficients for four quark operators in~Sec.~\ref{sec:fourq}, and 
those for quark bilinear operators in Sec.~\ref{sec:bilinear}.
Our final result of the matching coefficient for the kaon bag parameter is given in~Sec.~\ref{sec:conclusions}.

\section{Formulation}
\label{sec:formulation}

\subsection{Gradient flow}

In this section, we introduce the gradient flow with the background field method~\cite{Tec_02}, which simplifies perturbative calculations of renormalization factors. 
Our conventions for the gauge group factors and Casimirs are as follows:
We normalize the gauge group generators by
\begin{eqnarray}
\mathrm{Tr}(T^a T^b)=-T\,{\delta}^{ab}, \hspace{5mm}
[T^a,T^b]=f^{abc}T^c,
\end{eqnarray}
where $f^{abc}$ is the structure constant.
The anti-Hermitian matrices $T^a$ satisfy
\begin{eqnarray}
T^a T^a&=&-C_F{\1}.
\end{eqnarray}
For the fundamental representation of SU($N$), $T=1/2$, $\mathrm{dim}(R)=N$, and $C_F=(N^2-1)/2N$.

We first decompose the gauge field~$A_\mu$ and quark field~$\psi$ into background fields and quantum fields as
\begin{eqnarray}
A_\mu(x)&=&\hat{A}_\mu(x)+a_\mu(x),\\
\psi_f(x)&=&\hat{\psi}_f(x)+p_f(x),\\
\overline{\psi}_f(x)&=&\hat{\overline{\psi}}_f(x)+\overline{p}_f(x),\label{background_field}
\end{eqnarray} 
where $f=1$, 2, $\cdots$, $N_f$ is for the flavor, $\hat{A}_\mu$, $\hat{\psi}_f$, $\hat{\overline{\psi}}_f$ are background fields, and $a_\mu$, $p_f$, $\overline{p}_f$ are their quantum fields, respectively.

The flow equations we adopt are basically the simplest ones as proposed by L\"uscher~\cite{GF_04,GF_02}.  
The gradient flow drives the fields, $\hat{A}_\mu$, $\hat{\psi}_f$, $\hat{\overline{\psi}}_f$, $a_\mu$, $p_f$, and $\overline{p}_f$, into their flowed fields, $\hat{B}_\mu$, $\hat{\chi}_f,\,\hat{\overline{\chi}}_f$, $b_\mu$, $k_f$, and $\overline{k}_f$, respectively~\cite{Tec_02}.
Flow equations for the background fields are given by 
\begin{eqnarray}
\partial_t \hat B_\mu(t,x)&=&\hat D_\nu \hat G_{\nu\mu}(t,x), \;\;\;\; \hat B_\mu(t=0,x)=\hat A_\mu(x), \label{flow_eq_gauge_back_2}\\
\partial_t \hat{\chi}_f(t,x)&=&\hat{D}^2\hat\chi_f(t,x), \;\;\;\;\; \hat\chi_f(t=0,x)=\hat\psi_f(x), \label{flow_eq_fermion_back_2}\\
\partial_t \hat{\overline\chi}_f(t,x)&=&\hat{\overline\chi}_f(t,x)\hat{\overleftarrow{D}}^2, \;\;\;\;\; \hat{\overline\chi}_f(t=0,x)=\hat{\overline\psi}_f(x).\label{flow_eq_antifermion_back_2}
\end{eqnarray}
In this paper, we set the gauge parameter~$\alpha_0$ in~Ref.~\cite{Tec_02} to unity, $\alpha_0=1$. Then, the flow  equations for the quantum fields are given by
\begin{eqnarray}
\partial_t b_\mu(t,x)&=&\hat{D}^2b_\mu(t,x)+2[\hat{G}_{\mu\nu}(t,x),b_\nu(t,x)]+\hat{R}_\mu(t,x), 
\hspace{5mm} \hat b_\mu(t=0,x)=\hat a_\mu(x), \label{flow_eq_gauge_2}\\
\partial_t {k_f}(t,x)&=&\left\{D^2-\hat{D}_\mu{b}_\mu(t,x)\right\}k_f(t,x)+\left\{2b_\mu(t,x)\hat{D}_\mu+b^2(t,x)\right\}\hat{\chi}_f(t,x),\nonumber\\
&&\hspace{50mm} k_f(t=0,x)=p_f(x), \label{flow_eq_fermion_2}\\
\partial_t {\overline{k}}_f(t,x)&=&\overline{k}_f(t,x)\left\{\overleftarrow{D}^2+\hat{D}_\mu{b}_\mu(t,x)\right\}+ \hat{\overline{\chi}}_f(t,x)\left\{-2\hat{\overleftarrow{D}}_{\mu}b_\mu(t,x)+b^2(t,x)\right\},\nonumber\\
&&\hspace{50mm} {\overline{k}}_f(t=0,x)={\overline{p}}_f(x),\label{flow_eq_antifermion_2}
\end{eqnarray}
where we define 
\begin{eqnarray}
\hat{G}_{\mu\nu}(t,x)&=&\partial_t\hat{B}_\nu(t,x)-\partial_t\hat{B}_\mu(t,x)+[\hat{B}_\mu(t,x),\,\hat{B}_\nu(t,x)],\\
\hat{D}_\mu&=&\partial_\mu+[\hat{B}_\mu(t,x),\,\cdot\,],\;\;\;\;(\text{for gauge fields})\\
\hat{D}_\mu&=&\partial_\mu+\hat{B}_\mu(t,x), \;\;\;\;\;\;\;\;\;\; (\text{for quark fields})\\
\hat{R}_\mu(t,x)&=&2[b_\nu(t,x),\,\hat{D}_\nu{b}_\mu(t,x)] -[b_\nu(t,x),\,\hat{D}_\mu{b}_\nu(t,x)] \nonumber\\
&\;&+\;\left[b_\nu(t,x),\,\left[b_\nu(t,x),\,{b}_\mu(t,x)\right]\right].
\end{eqnarray}

In this paper, we set the background gauge field to zero and the background quark fields to constant. 
Then, the solution of the flow equations for the background fields is given by
\begin{eqnarray}
\hat{B}(t,x)&=&\hat{A}(x)=0,\\
\hat{\chi}_f(t,x)&=&\hat{\psi}_f(x)=(\text{const.}),\\
\hat{\overline{\chi}}_f(t,x)&=&\hat{\overline{\psi}}_f(x)=(\text{const.}).
\end{eqnarray}
Taking the solution of the background fields into account, the flow equations for the quantum fields can be simplified as
\begin{eqnarray}
\partial_t b_\mu(t,x)&=&{\partial}^2b_\mu(t,x)+\hat{R}_\mu(t,x), 
\hspace{5mm} \hat b_\mu(t=0,x)=\hat a_\mu(x), \label{flow_eq_gauge_3}\\
\partial_t {k_f}(t,x)&=&\left\{D^2-{\partial}_\mu{b}_\mu(t,x)\right\}k_f(t,x)+\left\{2b_\mu(t,x){\partial}_\mu+b^2(t,x)\right\}\hat{\psi}_f(t,x),\nonumber\\
&&\hspace{50mm} k_f(t=0,x)=p_f(x), \label{flow_eq_fermion_3}\\
\partial_t {\overline{k}}_f(t,x)&=&\overline{k}_f(t,x)\left\{\overleftarrow{D}^2+{\partial}_\mu{b}_\mu(t,x)\right\}+ \hat{\overline{\psi}}_f(t,x)\left\{-2{\overleftarrow{\partial}}_{\mu}b_\mu(t,x)+b^2(t,x)\right\},\nonumber\\
&&\hspace{50mm} {\overline{k}}_f(t=0,x)={\overline{p}}_f(x),\label{flow_eq_antifermion_3}
\end{eqnarray}
with
\begin{eqnarray}
\hat{R}^a_\mu(t,x)
=2f^{abc}{b}^b_\nu(t,x)\partial_\nu{b}^c_\mu(t,x)-f^{abc}{b}^b_\nu(t,x)\partial_\mu{b}^c_\nu(t,x)+f^{abc}f^{cde}{b}^b_\nu(t,x){b}^d_\nu(t,x){b}^e_\mu(t,x).\nonumber\\
\end{eqnarray}
Formal solution of the flow equations for the quantum fields is given by
\begin{eqnarray}
b^a_\mu(t,x)&=&e^{t{\partial}^2}a^a_\mu(x)+\int^t_0\mathrm{d}s\:e^{(t-s){\partial}^2}\hat{R}^a_\mu(s,x),\\
k_f(t,x)&=&e^{t{\partial}^2}p_f(x)\nonumber\\
&\;&+\int^t_0\mathrm{d}s\:e^{(t-s){\partial}^2}\left\{2b_\mu(s,x){\partial}_\mu+b^2(s,x)\right\}\left\{e^{s{\partial}^2}\hat{\psi}_f(x)+k_f(s,x)\right\},\\
\overline{k}_f(t,x)&=&\overline{p}_f(x)\;e^{t{\overleftarrow{\partial}}^2} \nonumber\\
&\;&+\int^t_0\mathrm{d}s\left\{\hat{\overline{\psi}}_f(x)\;e^{s{\overleftarrow{\partial}}^2}+\overline{k}_f(s,x)\right\}\left\{-2{\overleftarrow{\partial}}_\mu{b}_\mu(s,x)+b^2(s,x)\right\}e^{(t-s){\overleftarrow{\partial}}^2}.
\end{eqnarray}

In one-loop calculations discussed in this paper, we can disregard the $\mathcal{O}(\varg_0^4)$ terms in the propagators.
Thus, the propagator between $b(t,\ell)$ and $b(s,\ell)$ can be simplified as
\begin{eqnarray}
G^{ab}_{\mu\nu}(t,s;\ell) &\sim& e^{-(t+s)\ell^2}G^{ab}_{\mu\nu}(\ell)  \;\;\;\;\;\text{(one loop)}
\label{flowed_gauge_prop},
\end{eqnarray}
where 
\begin{eqnarray}
b^{a}_{\mu}(t,\ell)=\int \mathrm{d}^{D}x\:b^{a}_{\mu}(t,x)e^{-i\ell{\cdot}x}
\end{eqnarray}
is the quantum gauge field in the momentum space and $G^{ab}_{\mu\nu}(\ell)$ is the gluon propagator at $t=0$ with momentum $\ell$,
\begin{eqnarray}
G^{ab}_{\mu\nu}(\ell)=\varg^2_0\frac{1}{\ell^2}\delta^{ab}\delta_{\mu\nu}.
\end{eqnarray}
Similarly, the solution for quantum quark fields~$k_f$ and~$\overline{k}_f$ can also be simplified as
\begin{eqnarray}
k_f(t,x)&\sim&e^{t\partial^2}p_f(x)+\int^{t}_{0}\mathrm{d}s\:e^{(t-s)\partial^2}\left(b^2(s,x)\hat\psi_f+2b_\mu(s,x)\partial_\mu e^{s\partial^2}p_f(x)\right),\label{flowed_fermion} \\
\overline{k}_f(t,x)&\sim&\overline{p}_f(x)\:e^{t\overleftarrow{\partial}^2} 
 +\int^{t}_{0}\mathrm{d}s\:\left(\hat{\overline\psi}_fb^2(s,x)-2\overline{p}_f(x)\:e^{s\overleftarrow{\partial}^2}\overleftarrow{\partial}_\mu{b}_\mu(s,x)\right)e^{(t-s)\overleftarrow{\partial}^2}, 
\label{flowed_antifermion}
\end{eqnarray}
in one-loop calculations.

Because quark masses and external momenta appear as $tm_0^2$ and $tp^2$ in the matching coefficients, their dependence appear in higher orders of the flow time~$t$. 
Here, we set all quark masses and all external momenta of the four quark operators to zero for simplicity.

\subsection{Dimensional reduction scheme}

In the calculation of four quark operators, we use the Fierz rearrangement to organize the spinor indices. 
Because the Fierz rearrangement is defined for $4\times4$ Hermitian matrices, we have to restrict the spinor indices in the operator to run in the four dimensional space-time. 
In the dimensional regularization using the $D=4-2\epsilon$ dimensional space-time, 
we thus impose that only the internal loop momenta are reduced to the $D=4-2\epsilon$ dimensional space-time, while the other Lorentz indices run in four dimensional Lorentz space-time. 
This procedure is called the dimensional reduction scheme~\cite{DRED}. 

We denote the gamma matrices in four dimension as $\gamma_\mu$, and the gamma matrices in $D$ dimensional space-time as $\overline{\gamma}_\mu$. 
Denoting the remaining part as $\tilde{\gamma}_\mu$, the four dimensional gamma matrices are decomposed as
\begin{eqnarray}
\gamma_\mu &=& \overline{\gamma}_\mu+\tilde{\gamma}_\mu,\\
{\overline{\gamma}}_{\mu} &=& 
\left\{
\begin{array}{cl}
{\gamma}_{\mu}  &  (1 \leq \mu \leq D), \\
0  &  (D<\mu \leq 4),   
\end{array}
\right.
\\
{\tilde{\gamma}}_{\mu} &=& 
\left\{
\begin{array}{cl}
0  &  (1 \leq \mu \leq D), \\
{\gamma}_{\mu}  &  (D<\mu \leq 4).  
\end{array}
\right.
\end{eqnarray}
The anticommutation relation between $\gamma_\mu$ and $\overline{\gamma}_\nu$ can be calculated as
\begin{eqnarray}
\left\{\gamma_\mu,\,\overline{\gamma}_\nu\right\}=\left\{\left(\overline{\gamma}_\mu+\tilde{\gamma}_\mu\right),\,\overline{\gamma}_\nu\right\}=2\overline{\delta}_{\mu\nu},
\end{eqnarray}
where the $\overline{\delta}_{\mu\nu}$ means the Kronecker delta in $D$ dimension. 
The other relations can be calculated similarly, e.g.,
\begin{eqnarray}
\overline{\gamma}_\mu{\gamma}_\nu\overline{\gamma}_\mu &=& -D{\gamma}_\nu+2\overline{\gamma}_\nu,\label{formula_reduction_1}\\
{\gamma}_\mu\overline{\gamma}_\nu{\gamma}_\mu &=& -2\overline{\gamma}_\nu.\label{formula_reduction_2}
\end{eqnarray}

Finally, we define the $\gamma_5$ matrix which anticommutes with all the gamma matrices in this scheme:
\begin{eqnarray}
\left\{\gamma_5,\,\gamma_\mu\right\}&=&0,\\
\left\{\gamma_5,\,\overline{\gamma}_\mu\right\}&=&0,\\
\left\{\gamma_5,\,\tilde{\gamma}_\mu\right\}&=&0.
\end{eqnarray}
We construct four quark operators with $\gamma_\mu$ and $\gamma_5$, but the internal quark propagators contain $\overline{\gamma}_\mu$ only.

\subsection{Quark field renormalization}

\begin{figure}[htbp]
\begin{center}
\includegraphics[width=16.0cm, bb= 0 0 462 484]{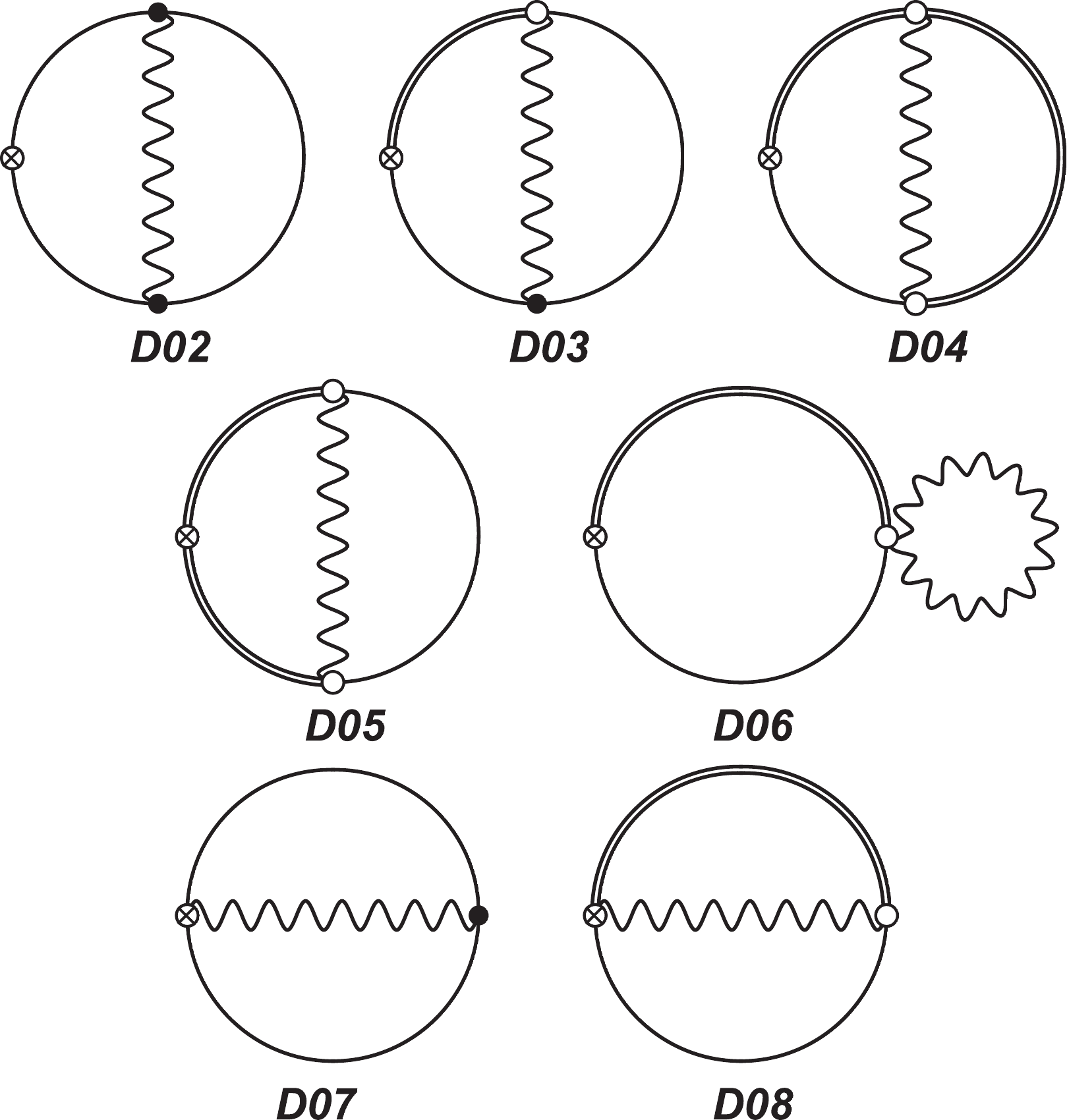}
\caption{One-loop Feynman diagrams for the quark field renormalization factor~$\varphi(t)$. See Ref.~\cite{GF_ap03} for the Feynman rule.}
\label{Fermion_ren_diag_2}
\end{center}
\end{figure}

It is known that, with the simple gradient flow driven by the pure gauge action as we adopt, quark field renormalization is required to keep the flowed fields finite~\cite{GF_02}.
Here, to avoid complications due to the matching between the lattice and dimensional regularization schemes, we adopt the quark field renormalization proposed in Ref.~\cite{GF_ap03}, in which the renormalized quark fields at $t>0$ (``ringed quark fields'') are given by
\begin{eqnarray}
\mathring{\chi}_f(t,x) &=& \sqrt{\frac{-2\:\text{dim}(R)}{(4\pi)^2t^2\left<\overline{\chi}_f(t,x)\gamma_\mu\overleftrightarrow{D}_\mu\chi_f(t,x)\right>}}\;\chi_f(t,x)
\; := \; \varphi^{1/2}(t)\;\chi_f(t,x).
\label{eq:varphi}
\end{eqnarray}
Note that the summation of the flavor index is not taken in this expression. Because we treat all quarks massless, the renormalization factor~$\varphi(t)$ is independent of $f$.

In~Ref~\cite{GF_ap03}, $\varphi(t)$ has been calculated to the one-loop order of the perturbation theory with the dimensional regularization scheme.
We revisit the calculation and compute~$\varphi(t)$ with the dimensional reduction scheme (DRED).
Feynman diagrams relevant to the quark field renormalization are listed in~Fig.~\ref{Fermion_ren_diag_2}. 
See~Ref.~\cite{GF_ap03} for the Feynman rule we adopt.
The diagrams mean
\begin{eqnarray}
D02&:&\int_{\ell,p} (-ip_\mu)e^{-2tp^2}S_{F\nu}(p)S_{F\rho}(p-\ell)S_{F\sigma}(p)G^{ab}_{\alpha\beta}(\ell)T^aT^b\text{tr}\left[\overline{\gamma}_\mu\overline{\gamma}_\nu\gamma_\alpha\overline{\gamma}_\rho\gamma_\beta\overline{\gamma}_\sigma\right],
\label{D02-1}\\
D03&:& \int^t_0\mathrm{d}s\int_{\ell,p}(-ip_\mu)e^{-(t-s)p^2}e^{-s(p-\ell)^2}e^{-tp^2}S_{F\nu}(p)S_{F\sigma}(p-\ell) 
\nonumber \\
&\:&\:\:\:\:\:\:\:\:\:\:\:\:\:\:\:\:\:\:\:\:\:\:\:\:\:\:\:\:\:\:\times\:
(-2i(p-\ell)_\lambda)G^{ab}_{\rho\lambda}(s,0;\ell)T^aT^b\text{tr}\left[\overline{\gamma}_\mu\overline{\gamma}_\nu{\gamma}_\rho\overline{\gamma}_\sigma\right], 
\label{D03-1}\\
D04&:& \int^t_0\mathrm{d}s\int^s_0\mathrm{d}u\int_{\ell,p}(-ip_\mu)S_{F\nu}(p)e^{-(t-s)p^2}e^{-(s-u)(p-\ell)^2}e^{-up^2}e^{-tp^2}
\nonumber\\
&\:&\:\:\:\:\:\:\:\:\:\:\:\:\:\:\:\:\:\:\:\:\:\:\:\:\:\:\:\:\:\:\times\:(-2i(p-\ell)_\rho)(-2ip_\lambda)G^{ab}_{\rho\lambda}(s,u;\ell)T^aT^b\text{tr}\left[\overline{\gamma}_\mu\overline{\gamma}_\nu\right],
\label{D04-1}\\
D05&:& \int^t_0\mathrm{d}s\int^t_0\mathrm{d}u\int_{\ell,p}(-ip_\mu)S_{F\nu}(p-\ell)e^{-(t-s)p^2}e^{-(t-u)p^2}e^{-(s+u)(p-\ell)^2}
\nonumber\\
&\:&\:\:\:\:\:\:\:\:\:\:\:\:\:\:\:\:\:\:\:\:\:\:\:\:\:\:\:\:\:\:\times\:(-2i(p-\ell)_\rho)(2i(p-\ell)_\sigma)G^{ab}_{\rho\sigma}(s,u;\ell)T^aT^b\text{tr}\left[\overline{\gamma}_\mu\overline{\gamma}_\nu\right],
\label{D05-1}\\
D06&:& \int^t_0\mathrm{d}s\int_{\ell,p}(-ip_\mu)S_{F\nu}(p)e^{-(t-s)p^2}e^{-sp^2}e^{-tp^2}G^{ab}_{\rho\rho}(s,s;\ell)T^aT^b\text{tr}\left[\overline{\gamma}_\mu\overline{\gamma}_\nu\right],
\label{D06-1}\\
D07&:&\int_{\ell,p} e^{-tp^2}e^{-t\ell^2}e^{-t(p+\ell)^2}S_{F\rho}(p)S_{F\lambda}(p+\ell)G^{ab}_{\mu\nu}(t,0;\ell)T^aT^b\text{tr}\left[{\gamma}_\mu\overline{\gamma}_\rho{\gamma}_\nu\overline{\gamma}_\lambda\right],
\label{D07-1}\\
D08&:& \int^t_0\mathrm{d}s\int_{\ell,p}\,e^{-(t-s)p^2}e^{-(t+s)(p+\ell)^2}(-2i(p-\ell)_\rho)S_{F\nu}(p+\ell)G^{ab}_{\mu\rho}(t,s;\ell)T^aT^b\text{tr}\left[{\gamma}_\mu\overline{\gamma}_\nu\right],
\nonumber\\
\label{D08-1}
\end{eqnarray}
where
$ \int_{\ell,p} = \int_\ell \int_p$ with $\int_\ell := \int d^D\ell / (2\pi)^D$, and
\begin{eqnarray}
S_{F\mu}(\ell)&=&-i\frac{\ell_\mu}{\ell^2},\\
G^{ab}_{\mu\nu}(t,s;\ell)&=&\varg^2_0e^{-(t+s)\ell^2}\frac{1}{\ell^2}\delta^{ab}\delta_{\mu\nu}.
\end{eqnarray}
Carrying out the computations similar to those given in~Ref.~\cite{GF_ap03}, 
we find that the diagrams contribute as 
\begin{eqnarray}
D02|_{\text{DRED}}&:&\;\; -\frac{1}{\epsilon}-2\log(8\pi t)-1,\\
D03|_{\text{DRED}}&:&\;\; 2\frac{1}{\epsilon}+4\log(8\pi t)+2+4\log(2)-2\log(3),\\
D04|_{\text{DRED}}&:&\;\; -20\log(2)+16\log(3),\\
D05|_{\text{DRED}}&:&\;\; 12\log(2)-5\log(3),\\
D06|_{\text{DRED}}&:&\;\; -4\frac{1}{\epsilon}-8\log(8\pi t)-4,\\
D07|_{\text{DRED}}&:&\;\; 8\log(2)-4\log(3),\\
D08|_{\text{DRED}}&:&\;\; -2\log(3),
\end{eqnarray}
in units of
\begin{eqnarray}
\frac{-2\:\text{dim}(R)}{(4\pi)^2t^2}\frac{\varg_0^2}{(4\pi)^2}\:C_F.
\end{eqnarray}

Collecting these contributions, we obtain the quark field renormalization factor in the dimensional reduction scheme, 
\begin{eqnarray}
\varphi(t)^\text{DRED}&=&(8\pi t)^{-\epsilon}\left\{1+\frac{\varg^2(\mu)}{(4\pi)^2}\:C_F\left(\frac{3}{\epsilon}+3\gamma_E+3\log{(2t\mu^2)+3-\log{(432)}}\right)\right\},
\label{Fermion_ren_t}
\end{eqnarray}
where we have replaced the bare gauge coupling~$\varg_0$ by the dimensionless~$\varg(\mu)$ using the prescription of the $\overline{\text{MS}}$ scheme~\cite{MS-bar},
\begin{eqnarray}
\varg_0^2=\left(\mu^{2}\frac{e^{\gamma_E}}{4\pi}\right)^{\epsilon}\varg^2(\mu) \left[ 1 + \mathcal{O}(\varg^2(\mu)) \right],
\label{eq:g0}
\end{eqnarray}
with $\gamma_E$ the Euler-Mascheroni constant.
Since $\varphi(t)$ is defined in terms of the expectation value of bare fields in Eq.~(\ref{eq:varphi}), it is independent of the renormalization scale~$\mu$. 
We may thus choose any value for~$\mu$ in the expressions above, 
provided that the perturbative expansions are well converged. 
Some conventional choices for $\mu$ are $\mu_d=1/\sqrt{8t}$~\cite{GF_04} and $\mu_0 \equiv 1/\sqrt{2e^{\gamma_E}t}$~\cite{EMT-2loop1}.

\section{Four quark operators}
\label{sec:fourq}

In this study, we consider four quark operators of the form
\begin{eqnarray}
O_{\pm}&=&\left[\left({\overline{\psi}}_{1} \gamma^{L}_{\mu}\psi_{2}\right)\left({\overline{\psi}}_{3} \gamma^{L}_{\mu}\psi_{4}\right)\pm\left({\overline{\psi}}_{1} \gamma^{L}_{\mu}\psi_{4}\right)\left({\overline{\psi}}_{3} \gamma^{L}_{\mu}\psi_{2}\right)\right], 
\label{def-op} 
\end{eqnarray}
where the subscripts $1,\cdots,4$ are for the flavor of the quark fields. 
We assume that these four flavors fulfill ${1}\neq{2}$, ${2}\neq{3}$, ${3}\neq{4}$, and ${4}\neq{1}$, to avoid closed quark loops within the four quark operator. 
For the calculation of $B_K$, the case $1=3\neq2=4$ is relevant.

We denote the background part of $O_{\pm}$ as
\begin{eqnarray}
\hat{O}_{\pm} &=& \left[\left(\hat{\overline{\psi}}_1 \gamma^{L}_{\mu}\hat\psi_2\right)\left(\hat{\overline{\psi}}_3 \gamma^{L}_{\mu}\hat\psi_4\right)\pm\left(\hat{\overline{\psi}}_1 \gamma^{L}_{\mu}\hat\psi_4\right)\left(\hat{\overline{\psi}}_3 \gamma^{L}_{\mu}\hat\psi_2\right)\right].
\end{eqnarray}
We then denote flowed four quark operators as
\begin{eqnarray}
O_{\pm}(t) &=& \left[\left({\overline{\chi}}_{1} \gamma^{L}_{\mu}\chi_{2}\right)\left({\overline{\chi}}_{3} \gamma^{L}_{\mu}\chi_{4}\right)\pm\left({\overline{\chi}}_{1} \gamma^{L}_{\mu}\chi_{4}\right)\left({\overline{\chi}}_{3} \gamma^{L}_{\mu}\chi_{2}\right)\right],
\label{def-op_t}
\end{eqnarray}
and their renormalized ones in terms of the ringed quark fields as 
\begin{eqnarray}
\mathring{O}_{\pm}(t)&=&\left[\left(\mathring{\overline{\chi}}_1 \gamma^{L}_{\mu}\mathring\chi_2\right)\left(\mathring{\overline{\chi}}_3 \gamma^{L}_{\mu}\mathring\chi_4\right)\pm\left(\mathring{\overline{\chi}}_1 \gamma^{L}_{\mu}\mathring\chi_4\right)\left(\mathring{\overline{\chi}}_3 \gamma^{L}_{\mu}\mathring\chi_2\right)\right].
\end{eqnarray}

Since the tree-level contributions of the flowed and bare operators are the same,
\begin{eqnarray}
\left<O_{\pm}(t)\right>_{\text{1PI}}|_{\text{tree}}&=&\hat{O}_{\pm} \;\;\;\;(\text{for }t\to0),\\
\left<O_{\pm}\right>_{\text{1PI}}|_{\text{tree}}&=&\hat{O}_{\pm},
\end{eqnarray}
we can write the small flow time expansion for $O_{\pm}(t)$ as
\begin{eqnarray}
O_{\pm}(t) &=& (1+I^{\text{GF}}_{\pm}(t)) \; O_{\pm}+\mathcal{O}(t),
\label{eq:IGF}
\end{eqnarray}
where we put the vertex correction as $I^{\text{GF}}_{\pm}(t)$.
To compute the $I^{\text{GF}}_{\pm}(t)$, it is convenient to consider one-particle irreducible vertex correction of $O_{\pm}(t)-O_{\pm}$, because, with the background field method, the vertex correction is proportional to the background part of the operator:
\begin{eqnarray}
\left<O_{\pm}(t)-O_{\pm}\right>_{\text{1PI}}&=&I^{\text{GF}}_{\pm}(t)\;\left<O_{\pm}\right>_{\text{1PI}}\nonumber\\
&=&I^{\text{GF}}_{\pm}(t)\; Z_{O_{\pm}}^{-1}\hat{O}_{\pm}
\;\sim\; I^{\text{GF}}_{\pm}(t)\;\hat{O}_{\pm}   \;\;\;\;\;\text{(one loop)}.
\label{4-2}
\end{eqnarray}
In the second line of~Eq.~(\ref{4-2}), we used the fact that the vertex correction $I^{\text{GF}}_{\pm}(t)$ is $\mathcal{O}(\varg_0^2)$.
From these relations, renormalized operators at small-$t$ are then given by
\begin{eqnarray}
\mathring{O}_{\pm}(t) &=& {(1+I^{\text{GF}}_{\pm}(t))}\;\left(\varphi^{\text{DRED}}(t)\right)^2 \; O_{\pm} +\mathcal{O}(t) .
\end{eqnarray}

We now consider renormalized four quark operators in the $\overline{\text{MS}}$ scheme with the dimensional reduction,
\begin{eqnarray}
O^{\overline{\text{MS}};\text{DRED}}_{\pm} &=& Z^{\overline{\text{MS}};\text{DRED}}_{O_{\pm}} \left(Z^{\overline{\text{MS}};\text{DRED}}_{\psi}\right)^2 O_{\pm} \end{eqnarray}
where $Z^{\overline{\text{MS}};\text{DRED}}_{O_{\pm}}$ and $Z^{\overline{\text{MS}};\text{DRED}}_{\psi}$ are renormalization factors for the four quark operator $O_{\pm}$ and for the quark field~$\psi(x)$, respectively.
Combining these relations, we obtain the one-loop expression for the matching coefficient~$Z^{\text{GF}\to\overline{\text{MS}};\text{DRED}}(t)$ to compute $O^{\overline{\text{MS}};\text{DRED}}_{\pm}$ from $\mathring{O}_{\pm}(t)$ in the $t\to0$ limit:
\begin{eqnarray}
O^{\overline{\text{MS}};\text{DRED}}_{\pm} &=& \lim_{t\to0} \; Z^{\text{GF}\to\overline{\text{MS}};\text{DRED}}_{O_{\pm}}(t) \; \mathring{O}_{\pm}(t),
\\
Z^{\text{GF}\to\overline{\text{MS}};\text{DRED}}_{O_{\pm}}(t) &=& \frac{Z^{\overline{\text{MS}};\text{DRED}}_{O_{\pm}}}{(1+I^{\text{GF}}_{\pm}(t))}\left(\frac{Z^{\overline{\text{MS}};\text{DRED}}_{\psi}}{\varphi^\text{DRED}(t)}\right)^2 .
\label{eq:ZGFMS}
\end{eqnarray}
The renormalization factor~$\varphi^\text{DRED}(t)$ is given by~Eq.~(\ref{Fermion_ren_t}). 
We calculate $I^{\text{GF}}_{\pm}(t)$, $Z^{\overline{\text{MS}};\text{DRED}}_{O_{\pm}}$, and $Z^{\overline{\text{MS}};\text{DRED}}_{\psi}$ in the following subsections. 

\subsection{Calculation of $I^{\text{GF}}_{\pm}(t)$}

\begin{figure}[htbp]
\begin{center}
\includegraphics[width=14.0cm,bb=0 0 462 280]{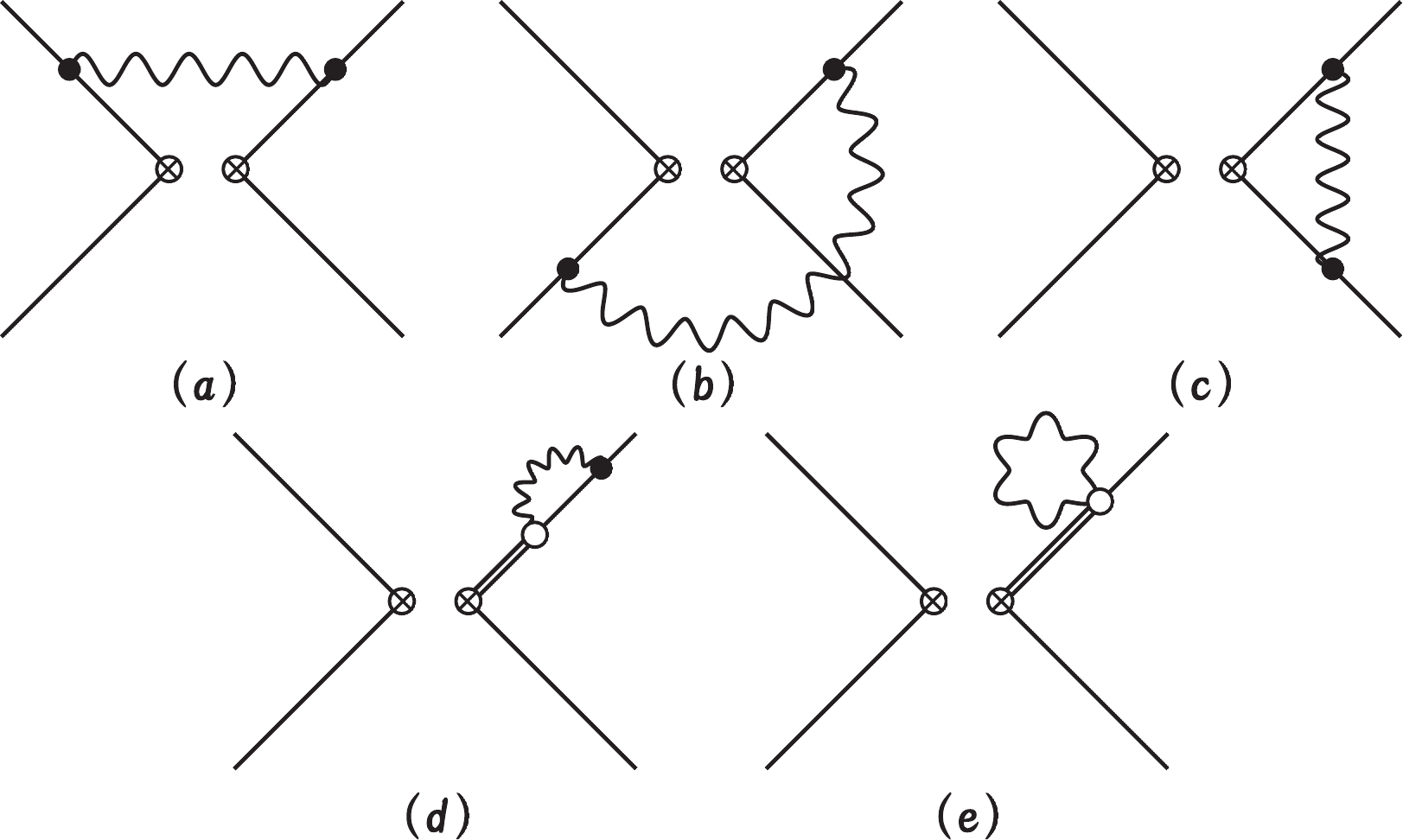}
\caption{One-loop 1PI diagrams for four quark operators with zero external momentum in the gradient flow scheme.}
\label{one_loop_diagram_S2}
\end{center}
\end{figure}

Setting all the external momenta to zero, we find that five diagrams shown in~Fig.~\ref{one_loop_diagram_S2} contribute to $ \left<O_{\pm}(t)-O_{\pm}\right>_{\text{1PI}} $ in the one-loop order.
Concrete forms of the diagrams are given by
\begin{eqnarray}
(a)&:& \int^{t}_{0}\mathrm{d}s \int_\ell ~\left[ \left(\hat{\overline{\psi}}_{1}~\gamma^L_\sigma\overline{\gamma}_{\rho}V^{a}_{\mu}~{\hat{\psi}}_{2}\right)\left(\hat{\overline{\psi}}_{3}~\gamma^L_\sigma\overline{\gamma}_{\lambda}V^{b}_{\nu}~{\hat{\psi}}_{4}\right) \pm\left\{\text{Fierz}\right\} \right] (-2\ell^2)e^{-2s\ell^2}S_{F\rho}(\ell)S_{F\lambda}(-\ell)G^{ab}_{\mu\nu}(\ell), 
\nonumber\\
\label{diagram-a}\\
(b)&:& \int^{t}_{0}\mathrm{d}s \int_\ell ~\left[ \left((\hat{\overline{\psi}}_{1}~V^{a}_{\mu}\overline{\gamma}_{\rho}\gamma^L_\sigma~{\hat{\psi}}_{2}\right)\left((\hat{\overline{\psi}}_{3}~\gamma^L_\sigma\overline{\gamma}_{\lambda}V^{b}_{\nu}~{\hat{\psi}}_{4}\right) \pm\left\{\text{Fierz}\right\} \right] (-2\ell^2)e^{-2s\ell^2}S_{F\rho}(\ell)S_{F\lambda}(\ell)G^{ab}_{\mu\nu}(\ell), 
\nonumber\\
\label{diagram-b}\\
(c)&:& \int^{t}_{0}\mathrm{d}s \int_\ell ~\left[ \left((\hat{\overline{\psi}}_{1}~V^{a}_{\mu}\overline{\gamma}_{\rho}\gamma^L_\sigma\overline{\gamma}_{\lambda}V^{b}_{\nu}~{\hat{\psi}}_{2}\right)\left((\hat{\overline{\psi}}_{3}~\gamma^L_\sigma~{\hat{\psi}}_{4}\right) \pm\left\{\text{Fierz}\right\} \right] (-2\ell^2)e^{-2s\ell^2}S_{F\rho}(\ell)S_{F\lambda}(\ell)G^{ab}_{\mu\nu}(\ell), 
\nonumber\\
\label{diagram-c}\\
(d)&:& 2 \int^{t}_{0}\mathrm{d}s \int_\ell ~\left[ \left((\hat{\overline{\psi}}_{1}~\gamma^L_\sigma({-i}){\ell_\mu}{T^a}\overline{\gamma}_\nu{V}^{b}_{\rho}~{\hat{\psi}}_{2}\right)\left((\hat{\overline{\psi}}_{3}~\gamma^L_\sigma~{\hat{\psi}}_{4}\right) \pm\left\{\text{Fierz}\right\} \right] e^{-s\ell^2}S_{F\nu}(\ell)G^{ab}_{\mu\rho}(s.0;\ell), 
\nonumber\\
\label{diagram-d} \\
(e)&:& \int^{t}_{0}\mathrm{d}s \int_\ell ~\left[ \left((\hat{\overline{\psi}}_{1}~\gamma^L_\sigma{T^a}{T^b}~{\hat{\psi}}_{2}\right)\left((\hat{\overline{\psi}}_{3}~\gamma^L_\sigma~{\hat{\psi}}_{4}\right) \pm\left\{\text{Fierz}\right\} \right] G^{ab}_{\mu\mu}(s,s;\ell), 
\label{diagram-e} 
\end{eqnarray}
where  
\begin{eqnarray}
V^{a}_{\mu} &=& \gamma_\mu T^a
\end{eqnarray}
is the quark-gluon vertex, and the symbol $\left\{\text{Fierz}\right\}$ means the Fierz partner of each original operator, \textit{i.e.},
\begin{eqnarray}
\left(\hat{\overline{\psi}}_{1}V^A{\hat{\psi}}_{2}\right)\left(\hat{\overline{\psi}}_{3}V^B{\hat{\psi}}_{4}\right)\pm\left\{\text{Fierz}\right\} &:=& \left(\hat{\overline{\psi}}_{1}V^A~{\hat{\psi}}_{2}\right)\left(\hat{\overline{\psi}}_{3}V^B{\hat{\psi}}_{4}\right)\pm\left(\hat{\overline{\psi}}_{1}V^A{\hat{\psi}}_{4}\right)\left(\hat{\overline{\psi}}_{3}V^B{\hat{\psi}}_{2}\right).
\nonumber\\
\end{eqnarray}
with $V^A$ and $V^B$ some combinations of $\gamma_\mu$, $T^a$, etc.

\subsubsection{Contribution of diagrams $c$, $d$ and $e$}

We first evaluate the diagrams $(c)$, $(d)$ and $(e)$ of~Fig.~\ref{one_loop_diagram_S2}. 
The calculation is similar to those for fermion bilinear operators discussed in~Ref.~\cite{Tec_04}.
The main difference comes from $\mathcal{O}(\epsilon)$ term called the evanescent operator, 
which is a byproduct of the dimensional reduction scheme~\cite{Evanescent01,Evanescent02}.

The spinor factor of the diagram $(c)$ is calculated as
\begin{eqnarray}
\left(\gamma_\mu \overline{\gamma}_\rho \gamma^L_\sigma \overline{\gamma}_\rho \gamma_{\mu}\right)_{\alpha\beta}\left(\gamma^L_\sigma\right)_{\gamma\delta}=2D\left(\gamma^L_\sigma\right)_{\alpha\beta}\left(\gamma^L_\sigma\right)_{\gamma\delta} - 4\left(\overline{\gamma}^L_\sigma\right)_{\alpha\beta}\left(\overline{\gamma}^L_\sigma\right)_{\gamma\delta}.
\label{eq:diagramc-1}
\end{eqnarray}
The new Dirac structure $(\overline{\gamma}^L_\sigma)_{\alpha\beta}(\overline{\gamma}^L_\sigma)_{\gamma\delta}$ must be removed appropriately to achieve the correct physical operator. Then, we define the corresponding evanescent operator~$\hat{E}$~\cite{Evanescent01,Evanescent02} by
\begin{eqnarray}
\hat{E} &:=& \left(\gamma^L_\sigma\right)_{\alpha\beta}\left(\gamma^L_\sigma\right)_{\gamma\delta} - \frac{4}{D}\left(\overline{\gamma}^L_\sigma\right)_{\alpha\beta}\left(\overline{\gamma}^L_\sigma\right)_{\gamma\delta}.\label{evanescent_op} \\
&=& \left(\tilde{\gamma}^L_\sigma\right)_{\alpha\beta}\left(\tilde{\gamma}^L_\sigma\right)_{\gamma\delta} - \frac{\epsilon}{2}\left({\gamma}^L_\sigma\right)_{\alpha\beta}\left({\gamma}^L_\sigma\right)_{\gamma\delta}+\mathcal{O}(\epsilon^2).
\label{eq:ev2}
\end{eqnarray}
Because the remnant gamma matrices $\tilde{\gamma}_\mu$ live in the $2\epsilon$ dimensional space, we consider that the first term of~Eq.~(\ref{eq:ev2}) is $\mathcal{O}(\epsilon)$ and thus $\hat{E}$ itself is $\mathcal{O}(\epsilon)$.
For the diagram~$(c)$ of~Fig.~\ref{one_loop_diagram_S2}, we subtract~$D \hat{E}$ from Eq.~(\ref{eq:diagramc-1}) to obtain
\begin{eqnarray}
\left(\gamma_\mu \overline{\gamma}_\rho \gamma^L_\sigma \overline{\gamma}_\rho \gamma_{\mu}\right)_{\alpha\beta}\left(\gamma^L_\sigma\right)_{\gamma\delta}=D\left(\gamma^L_\sigma\right)_{\alpha\beta}\left(\gamma^L_\sigma\right)_{\gamma\delta}.
\end{eqnarray}
Note that the definition of evanescent operator links to a finite renormalization (or subtraction) of four quark operators because of $\mathcal{O}(1/\epsilon)$ UV divergences.
Together with its Fierz partner, $\hat{O}_\pm$ is formed. 
The spinor factors for other diagrams can be calculated similarly.

The integrations over the internal momentum can be evaluated by the formula,
\begin{eqnarray}
\int_\ell \frac{1}{\ell^2}\:e^{-t\ell^2}=\frac{t^{1-D/2}}{(4\pi)^{D/2}}\frac{\Gamma(D/2-1)}{\Gamma(D/2)},
\end{eqnarray}
where $\Gamma(x)$ is the gamma function. 
We find that Eqs.~(\ref{diagram-c})$\sim$(\ref{diagram-e}) are evaluated as
\begin{eqnarray}
(c)&:&\;\; \frac{-\varg_0^2}{(4\pi)^2}C_F\left\{\frac{1}{\epsilon}+\log(8\pi{t})+1\right\}\hat{O}_{\pm}, \label{result-c} \\
(d)&:&\;\; \frac{\varg_0^2}{(4\pi)^2}C_F\left\{\frac{1}{\epsilon}+\log(8\pi{t})+1\right\}\hat{O}_{\pm}, \label{result-d} \\
(e)&:&\;\; \frac{-2\varg_0^2}{(4\pi)^2}C_F\left\{\frac{1}{\epsilon}+\log(8\pi{t})+1\right\}\hat{O}_{\pm}. \label{result-e} 
\end{eqnarray}

\subsubsection{Contribution of diagrams $a$ and $b$}

We now evaluate the diagrams $(a)$ and $(b)$.
The color indices in~Eqs.~(\ref{diagram-a})  and (\ref{diagram-b}) can be handled using the relation
\begin{eqnarray}
T^a_{ij}T^a_{kl}=- T\left(\delta_{il}\delta_{jk}-\frac{1}{\text{dim}(R)}\delta_{ij}\delta_{kl}\right).
\end{eqnarray}
The complicated structure of the spinor indices in~Eqs.~(\ref{diagram-a}) and (\ref{diagram-b}) can be simplified using the Fierz rearrangement,
\begin{eqnarray}
\left(\Lambda_{1}\right)_{\alpha\beta}\left(\Lambda_{2}\right)_{\gamma\delta}
&=&-\frac{1}{4}\sum_{\Gamma^A}\left(\Lambda_{1}\Gamma^A\Lambda_{2}\right)_{\alpha\delta}\left(\Gamma^A\right)_{\gamma\beta}\\
\Gamma^A&=&\left\{\1,\,\gamma_5,\,\gamma_\mu,\,i\gamma_\mu\gamma_5,\sigma_{\mu\nu}\right\},
\end{eqnarray}
where $\sigma_{\mu\nu} = \frac{i}{2}[\gamma_\mu,\,\gamma_\nu]$.
Using relations (\ref{formula_reduction_1}) and (\ref{formula_reduction_2}) of the dimensional reduction scheme, we find
\begin{eqnarray}
(a): \: \left(\gamma^L_\sigma \overline{\gamma}_\rho \gamma_{\mu}\right)_{\alpha\beta}\left(\gamma^L_\sigma \overline{\gamma}_\rho \gamma_{\mu}\right)_{\gamma\delta}&=&4\,D\left(\gamma^L_\sigma\right)_{\alpha\delta}\left(\gamma^L_\sigma\right)_{\gamma\beta}
\;=\;4\,D\left(\gamma^L_\sigma\right)_{\alpha\beta}\left(\gamma^L_\sigma\right)_{\gamma\delta}, \label{spinor-a} \\
(b): \: \left(\gamma_{\mu} \overline{\gamma}_\rho \gamma^L_\sigma\right)_{\alpha\beta}\left(\gamma^L_\sigma \overline{\gamma}_\rho \gamma_{\mu}\right)_{\gamma\delta}&=&2\,D\left(\gamma^L_\sigma\right)_{\alpha\delta}\left(\gamma^L_\sigma\right)_{\gamma\beta}-4\left(\overline{\gamma}^L_\sigma\right)_{\alpha\delta}\left(\overline{\gamma}^L_\sigma\right)_{\gamma\beta}\nonumber\\
&=&D\left(\gamma^L_\sigma\right)_{\alpha\beta}\left(\gamma^L_\sigma\right)_{\gamma\delta}-D\hat{E}
\;=\;D\left(\gamma^L_\sigma\right)_{\alpha\beta}\left(\gamma^L_\sigma\right)_{\gamma\delta}, \label{spinor-b}
\end{eqnarray}
where the evanescent operator defined by Eq.~(\ref{evanescent_op}) is removed to obtain the second line of~Eq.~(\ref{spinor-b}).

Carrying out the integrations, we obtain the contributions of the diagrams $(a)$ and $(b)$ given by
\begin{eqnarray}
(a)&:&\;\; \frac{-4\varg_0^2}{(4\pi)^2}\left(\frac{T}{\text{dim}(R)}\mp{T}\right)\left\{\frac{1}{\epsilon}+\log(8\pi{t})+1\right\}\hat{O}_{\pm}, \label{result-a} \\
(b)&:&\;\; \frac{\varg_0^2}{(4\pi)^2}\left(\frac{T}{\text{dim}(R)}\mp{T}\right)\left\{\frac{1}{\epsilon}+\log(8\pi{t})+1\right\}\hat{O}_{\pm}. \label{result-b} 
\end{eqnarray}

\subsubsection{Result for $I^{\text{GF}}_{\pm}(t)$}

We now combine the results of~Eqs.~(\ref{result-c}), (\ref{result-d}), (\ref{result-e}), (\ref{result-a}), and~(\ref{result-b}), 
taking into account the fact that there exist two different diagrams for each of the types~$(a)$, $(b)$, and~$(c)$, while four diagrams for each of the types~$(d)$ and~$(e)$. 
Note that, by removing the evanescent operator defined by Eq.~(\ref{evanescent_op}), the background fields are correctly combined to form the $\hat{O}_\pm$.
Our result for the coefficient $I^{\text{GF}}_{\pm}(t)$ in front of the $\hat{O}_\pm$ is given by 
\begin{eqnarray}
I^{\text{GF}}_{\pm}(t) &=& -6\frac{\varg^2(\mu)}{(4\pi)^2}\left(\frac{T}{\text{dim}(R)}\mp{T}+C_F\right)\left\{\frac{1}{\epsilon}+\gamma_E+\log(2t\mu^2)+1\right\},
\label{eq:IGFfinal}
\end{eqnarray}
where we have replaced $\varg_0$ by $\varg(\mu)$ using~Eq.~(\ref{eq:g0}).
From the definition of $I_\pm^{\text{GF}}$ given in~Eq.~(\ref{eq:IGF}), 
$I_\pm^{\text{GF}}$ is independent of the renormalization scale~$\mu$. 
We may choose any value for~$\mu$ provided that the perturbative expansions are well converged.

\subsection{$\overline{\text{MS}}$ renormalization factors $Z^{\overline{\text{MS}};{\text{DRED}}}_{\psi}$ and $Z^{\overline{\text{MS}};{\text{DRED}}}_{O_{\pm}}$}

\begin{figure}[tbh]
\begin{center}
\includegraphics[width=14.0cm,bb=0 0 462 138]{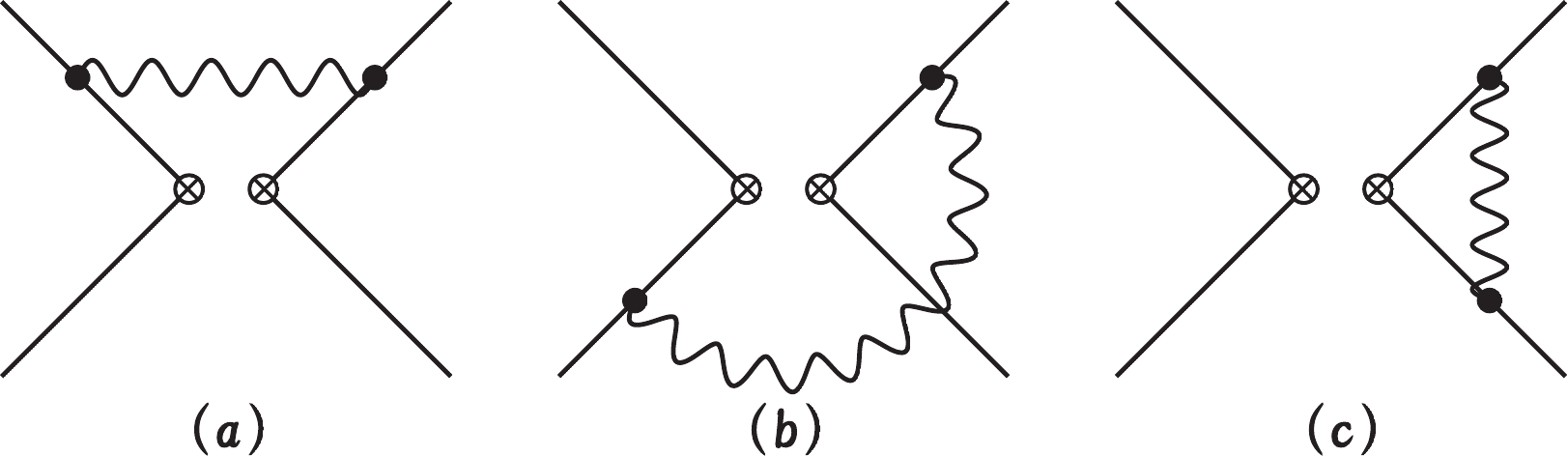}
\caption{One-loop 1PI diagrams for four quark operators with zero external momentum in the $\overline{\text{MS}}$ scheme.}
\label{one_loop_diagram_S2_MS}
\end{center}
\end{figure}

The last pieces to be calculated are the renormalization factors for the quark field~$\psi(x)$ and the four quark operator~$O_{\pm}$ in the $\overline{\text{MS}}$ scheme with the dimensional reduction.
See Refs.~\cite{Tec_05,Tec_07} for previous efforts to connect four quark operators in the $\overline{\text{MS}}$ scheme with those in the lattice scheme.
We again set the external momentum to zero, but introduce a gluon mass~$\lambda$ to regularize the infrared divergences. 
Then, the gluon propagator~$G^{ab}_{\mu\nu}(\ell)$ is given by
\begin{eqnarray}
G^{ab}_{\mu\nu}(\ell;\lambda)&=&\varg_0^2\frac{1}{\ell^2+\lambda^2}\delta^{ab}\delta_{\mu\nu}.
\end{eqnarray}
A convenient formula in these calculations is
\begin{eqnarray}
\int_\ell \frac{1}{(\ell^2)^a(\ell^2+\lambda^2)}=\frac{1}{(4\pi)^{D/2}}\lambda^{D-2a-2}\frac{\Gamma(D/2-a)\Gamma(a+1-D/2)}{\Gamma(D/2)}.\label{formula00}
\end{eqnarray}

The renormalization factor~$Z^{\overline{\text{MS}};{\text{DRED}}}_{\psi}$ for the quark field is calculated via the self-energy.
Denoting $\slashed{p} = p_\mu \overline{\gamma}_\mu$, we find
\begin{eqnarray}
\left<\psi(x)\overline\psi(y)\right>
&=&\int_p\,\frac{1}{i\slashed{p}}e^{ip\cdot(x-y)}- \varg_0^2 ~C_F \int_{p,q}\,\frac{1}{i\slashed{p}}\gamma_{\mu}\frac{1}{i\slashed{q}}\gamma_{\mu}\frac{1}{i\slashed{p}}\,\frac{1}{(p-q)^2+\lambda^2}e^{ip\cdot(x-y)},
\nonumber
\\
&=&\int_p \frac{1}{i\slashed{p}} \left[ 1  
- \frac{\varg_0^2}{(4\pi)^2} C_F \left\{\frac{1}{\epsilon} - \gamma_E + \log\left(\frac{4\pi}{\lambda^2}\right)+\frac{1}{2}\right\} \right] e^{ip\cdot(x-y)}
\nonumber
\\
&=&\int_p \frac{1}{i\slashed{p}} \left[ 1  
-\frac{\varg^2(\Tilde\mu)}{(4\pi)^2} C_F \left\{\frac{1}{\epsilon}+\log\left(\frac{{\Tilde\mu}^2}{\lambda^2}\right)+\frac{1}{2}\right\} \right] e^{ip\cdot(x-y)}
\label{eq:psipsi}
\end{eqnarray}
in the dimensional reduction scheme.
In the last line of Eq.~(\ref{eq:psipsi}), we have replaced $\varg_0$ by $\varg(\Tilde\mu)$ using~Eq.~(\ref{eq:g0}), 
where $\Tilde\mu$ is the renormalization scale for the $\overline{\text{MS}}$ scheme. 
A conventional choice for $\Tilde\mu$ is 2~GeV.
Thus the quark field renormalization factor reads
\begin{eqnarray}
Z^{\overline{\text{MS}};{\text{DRED}}}_{\psi}=1+\frac{\varg^2(\Tilde\mu)}{(4\pi)^2} C_F \,\frac{1}{\epsilon}
\label{eq:ZMSpsi}
\end{eqnarray}
in the one-loop order of the dimensional reduction scheme.

The renormalization factors~$Z^{\overline{\text{MS}};{\text{DRED}}}_{O_{\pm}}$ for the four quark operators~$O_\pm$ are evaluated considering the 1PI vertex corrections to $O_{\pm}$. 
Diagrams showing what we need to evaluate are given in~Fig.~\ref{one_loop_diagram_S2_MS}. 
The contributions from the three diagrams are given by
\begin{eqnarray}
(a)&:&\;\;  \int_\ell \left[ \left(\hat{\overline{\psi}}_{1}~\gamma^L_\sigma\overline{\gamma}_{\rho}V^{a}_{\mu}~{\hat{\psi}}_{2}\right)\left(\hat{\overline{\psi}}_{3}~\gamma^L_\sigma\overline{\gamma}_{\lambda}V^{b}_{\nu}~{\hat{\psi}}_{4}\right) \pm\left\{\text{Fierz}\right\} \right] S_{F\rho}(\ell)S_{F\lambda}(-\ell)G^{ab}_{\mu\nu}(\ell;\lambda), 
\label{diagram-a-normal}\\
(b)&:&\;\;  \int_\ell \left[ \left(\hat{\overline{\psi}}_{1}~V^{a}_{\mu}\overline{\gamma}_{\rho}\gamma^L_\sigma~{\hat{\psi}}_{2}\right)\left(\hat{\overline{\psi}}_{3}~\gamma^L_\sigma\overline{\gamma}_{\lambda}V^{b}_{\nu}~{\hat{\psi}}_{4}\right) \pm\left\{\text{Fierz}\right\} \right] S_{F\rho}(\ell)S_{F\lambda}(\ell)G^{ab}_{\mu\nu}(\ell;\lambda), 
\label{diagram-b-normal}\\
(c)&:&\;\;  \int_\ell \left[ \left(\hat{\overline{\psi}}_{1}~V^{a}_{\mu}\overline{\gamma}_{\rho}\gamma^L_\sigma\overline{\gamma}_{\lambda}V^{b}_{\nu}~{\hat{\psi}}_{2}\right)\left(\hat{\overline{\psi}}_{3}~\gamma^L_\sigma~{\hat{\psi}}_{4}\right) \pm\left\{\text{Fierz}\right\} \right] S_{F\rho}(\ell)S_{F\lambda}(\ell)G^{ab}_{\mu\nu}(\ell;\lambda). 
\label{diagram-c-normal}
\end{eqnarray}
We again use the Fierz rearrangement adopting the same evanescent operator defined by~Eq.~(\ref{evanescent_op}). 
We obtain
\begin{eqnarray}
(a)&:&\;\; \frac{4\varg_0^2}{(4\pi)^2} \left(\frac{T}{\text{dim}(R)}\mp{T}\right) \left\{\frac{1}{\epsilon}-\gamma_E +\log\left(\frac{4\pi}{\lambda^2}\right)+1\right\}\hat{O}_{\pm}, \label{result-a-MS}\\
(b)&:&\;\; \frac{-\varg_0^2}{(4\pi)^2} \left(\frac{T}{\text{dim}(R)}\mp{T}\right) \left\{\frac{1}{\epsilon}-\gamma_E+\log\left(\frac{4\pi}{\lambda^2}\right)+1\right\}\hat{O}_{\pm}, \label{result-b-MS}\\
(c)&:&\;\; \frac{\varg_0^2}{(4\pi)^2} C_F \left\{\frac{1}{\epsilon}-\gamma_E +\log\left(\frac{4\pi}{\lambda^2}\right)+1\right\}\hat{O}_{\pm}. \label{result-c-MS}
\end{eqnarray}
Collecting them, we obtain the one-loop 1PI vertex correction 
\begin{eqnarray}
\left<{O}_{\pm}\right>_{\text{1PI}}=\left[1+2\frac{\varg^2(\Tilde\mu)}{(4\pi)^2} \left(\frac{3T}{\text{dim}(R)}\mp{3T}+C_F\right) \left\{\frac{1}{\epsilon}+\log{\left(\frac{{\Tilde\mu}^2}{\lambda^2}\right)}+1\right\}\right]\hat{O}_{\pm}.
\end{eqnarray}
The $\overline{\text{MS}}$ renormalization factor in the dimensional reduction scheme is extracted as
\begin{eqnarray}
Z^{\overline{\text{MS}};{\text{DRED}}}_{O_{\pm}}=1-2\frac{\varg^2(\Tilde\mu)}{(4\pi)^2}\left(\frac{3T}{\text{dim}(R)}\mp{3T}+C_F\right)\,\frac{1}{\epsilon}.
\label{eq:ZMSO}
\end{eqnarray}

\subsection{Matching coefficient for four quark operators $O_\pm$}

Combining the results of~Eqs.~(\ref{Fermion_ren_t}), (\ref{eq:IGFfinal}), (\ref{eq:ZMSpsi}), and~(\ref{eq:ZMSO}) for $\varphi^{\text{DRED}}(t)$, $I^{\text{GF}}_{\pm}(t)$, $Z^{\overline{\text{MS}};{\text{DRED}}}_{\psi}$, and $Z^{\overline{\text{MS}};{\text{DRED}}}_{O_{\pm}}$,
we find that the matching coefficient for $O_\pm$ is given by
\begin{eqnarray}
Z^{\text{GF}\to\overline{\text{MS}};\text{DRED}}_{O_{\pm}}(t)
&=&\frac{Z^{\overline{\text{MS}};{\text{DRED}}}_{O_{\pm}}}{(1+I^{\text{GF}}_{\pm}(t))}\left(\frac{Z^{\overline{\text{MS}};{\text{DRED}}}_{\psi}}{\varphi^{\text{DRED}}(t)}\right)^2\nonumber\\
&=&1
+6\,\frac{\varg^2(\mu)-\varg^2(\Tilde{\mu})}{(4\pi)^2}
\left(\frac{T}{\text{dim}(R)}\mp{T}\right)
\frac{1}{\epsilon}
\label{matching_factor0}\\
&&
+\frac{\varg^2(\mu)
}{(4\pi)^2}\,\left\{6\left(\frac{T}{\text{dim}(R)}\mp{T}\right)\left(\log{(2t\mu^2)}+\gamma_E+1\right)+2C_F\log{432}\right\}.\nonumber
\end{eqnarray}
From~Eq.~(\ref{eq:g0}), we have the tree-level running of the coupling constant,
\begin{equation}
   \mu\frac{d}{d\mu}\varg^2(\mu)=-2\epsilon \varg^2(\mu).
\end{equation}
Integrating this equation, we obtain
\begin{eqnarray}
   \varg^2(\mu)-\varg^2(\Tilde{\mu})
   &=&\left[1-\left(\frac{\Tilde{\mu}^2}{\mu^2}\right)^{-\epsilon}\right]
   \varg^2(\mu)
\nonumber\\
   &=&\epsilon\log\left(\frac{\Tilde{\mu}^2}{\mu^2}\right) \varg^2(\mu)
   +O(\epsilon^2).
\end{eqnarray}
We thus find that the one-loop matching coefficient for $O_{\pm}(t)$ is given by
\begin{eqnarray}
Z^{\text{GF}\to\overline{\text{MS}};\text{DRED}}_{O_{\pm}}(t)
&=&1
+\frac{\varg^2(\mu)}{(4\pi)^2}\,\left\{6\left(\frac{T}{\text{dim}(R)}\mp{T}\right)
\left(\log{(2t{\Tilde\mu}^2)}+\gamma_E +1\right)+2C_F\log{432}\right\} ,
\nonumber\\
\label{eq:matching_coefficient_Opm}
\end{eqnarray}
where $\mu^2$ in the $\log$ of~Eq.~(\ref{matching_factor0}) is replaced by ${\Tilde\mu}^2$ due to the contribution from $\varg^2(\mu)-\varg^2(\Tilde{\mu})$.

With the matching coefficient $Z^{\text{GF}\to\overline{\text{MS}};{\text{DRED}}}_{O_{\pm}}(t)$, we evaluate the $\overline{\text{MS}}$ renormalized four quark operators~$O^{\overline{\text{MS}};\text{DRED}}_{\pm}$ in the dimensional reduction scheme from the corresponding lattice operators~$\mathring{O}_{\pm}(t)$ at small flow time~$t$, 
\begin{eqnarray}
O^{\overline{\text{MS}};\text{DRED}}_{\pm} &=& \lim_{t\to0} \; Z^{\text{GF}\to\overline{\text{MS}};{\text{DRED}}}_{O_{\pm}}(t) \; \mathring{O}_{\pm}(t).
\label{eq:matchingOpm2}
\end{eqnarray}
Note that the $1/\epsilon$ UV divergences in~Eqs.~(\ref{Fermion_ren_t}), (\ref{eq:IGFfinal}), (\ref{eq:ZMSpsi}), and~(\ref{eq:ZMSO}) cancel out with each other in the combination of the matching coefficient~$Z^{\text{GF}\to\overline{\text{MS}};{\text{DRED}}}_{O_{\pm}}(t)$.
This is expected from the finiteness of the matching coefficients in the SF\textit{t}X method:
Because both $O^{\overline{\text{MS}};{\text{DRED}}}_{\pm}$ and $\mathring{O}_{\pm}(t)$ are finite in the matching relation (\ref{eq:matchingOpm2}), $Z^{\text{GF}\to\overline{\text{MS}};{\text{DRED}}}_{O_{\pm}}(t)$ should also be finite.
This is explicitly confirmed by~Eq.~(\ref{eq:matching_coefficient_Opm}).

\section{Quark bilinear operators}
\label{sec:bilinear}

To calculate the kaon bag parameter, we also need the matching coefficient of the quark bilinear operator in the denominator of Eq.~(\ref{eq:BK}).
In this study, we consider general bilinear operators of the form 
\begin{eqnarray}
\overline{\psi}_1 \,\Gamma\, \psi_2
\label{eq:bilinear}
\end{eqnarray}
with $\Gamma =\1$, $\gamma_5$, $\gamma_{\mu}$, $i\gamma_{\mu}\gamma_{5}$, and $\sigma_{\mu\nu}$. 
We assume that the flavors satisfy $1\neq2$ to avoid a closed quark loop within the operator.
The calculations in the dimensional regularization scheme are similar to those given in Sec.~\ref{sec:fourq}.
We thus just show the final results.

We find that, at small flow time $t$, the one-loop 1PI vertex corrections for the bilinear operators are given by
\begin{eqnarray}
\left<\overline{\chi}_1(t) \,\Gamma\, \chi_2(t)-\overline{\psi}_1 \,\Gamma\, \psi_2\right>_{\text{1PI}}&=&I^{\text{GF}}_{\Gamma}(t)\left(\hat{\overline{\psi}}_1 \,\Gamma\, \hat\psi_2 \right)
\end{eqnarray}
with
\begin{eqnarray}
I^{\text{GF}}_{\Gamma}(t)&=&
\left\{
\begin{array}{ll}
(-6)\frac{\varg^2(\mu)}{(4\pi)^2}C_F\left\{\frac{1}{\epsilon}+\gamma_E+\log(2t\mu^2)+1\right\} ,\:\: &\Gamma=\1,\,\gamma_5, \\
(-3)\frac{\varg^2(\mu)}{(4\pi)^2}C_F\left\{\frac{1}{\epsilon}+\gamma_E+\log(2t\mu^2)+1\right\} ,\:\: &\Gamma=\gamma_{\mu},\,i\gamma_{\mu}\gamma_{5}, \\
(-2)\frac{\varg^2(\mu)}{(4\pi)^2}C_F\left\{\frac{1}{\epsilon}+\gamma_E+\log(2t\mu^2)+1\right\} ,\:\: &\Gamma=\sigma_{\mu\nu},
\end{array}
\right.
\end{eqnarray}
where we have replaced $\varg_0$ by $\varg(\mu)$ using~Eq.~(\ref{eq:g0}).
The evanescent operators we adopt are defined by
\begin{eqnarray}
\hat{E}_{\Gamma}=
\left\{
\begin{array}{ll}
0 ,\:\: &\Gamma=\1,\,\gamma_5, \\
{\gamma}_{\mu}-\frac{4}{D} \overline{\gamma}_\mu ,\:\: &\Gamma=\gamma_\mu,\\
i{\gamma}_{\mu} \gamma_5-i\frac{4}{D} \overline{\gamma}_\mu \gamma_5 ,\:\: &\Gamma=i\gamma_\mu \gamma_5,\\
(D-4)\sigma_{\mu\nu} ,\:\: &\Gamma=\sigma_{\mu\nu}.
\end{array}
\right.
\end{eqnarray}

Corresponding results at $t=0$ are given by
\begin{eqnarray}
\left<\overline{\psi}_1 \,\Gamma\, \psi_2\right>_{\text{1PI}}&=&I^{\overline{\text{MS}};\text{DRED}}_{\Gamma}\left(\hat{\overline{\psi}}_1 \,\Gamma\, \hat\psi_2 \right),
\end{eqnarray}
with
\begin{eqnarray}
I^{\overline{\text{MS}};\text{DRED}}_{\Gamma}&=&
\left\{
\begin{array}{ll}
1+4\frac{\varg^2(\Tilde{\mu})}{(4\pi)^2}C_F\left\{\frac{1}{\epsilon}+\log\left(\frac{{\Tilde\mu}^2}{\lambda^2}\right)+1\right\} ,\:\: &\Gamma=\1,\,\gamma_5,\\
1+\frac{\varg^2(\Tilde{\mu})}{(4\pi)^2}C_F\left\{\frac{1}{\epsilon}+\log\left(\frac{{\Tilde\mu}^2}{\lambda^2}\right)+1\right\} ,\:\: &\Gamma=\gamma_{\mu},\,i\gamma_{\mu}\gamma_{5},\\
1 ,\:\: &\Gamma=\sigma_{\mu\nu},
\end{array}
\right.
\end{eqnarray}
where we have replaced $\varg_0$ by $\varg(\Tilde\mu)$ with setting the renormalization scale of the $\overline{\text{MS}}$ scheme to $\Tilde\mu$. 
From the results of $I^{\overline{\text{MS}};\text{DRED}}_{\Gamma}$, we obtain the $\overline{\text{MS}}$ renormalization factors, 
\begin{eqnarray}
Z^{\overline{\text{MS}};\text{DRED}}_{\Gamma}&=&
\left\{
\begin{array}{ll}
1-4\frac{\varg^2(\tilde\mu)}{(4\pi)^2}C_F\frac{1}{\epsilon} ,\:\: &\Gamma=\1,\,\gamma_5,\\
1-\frac{\varg^2(\tilde\mu)}{(4\pi)^2}C_F\frac{1}{\epsilon} ,\:\: &\Gamma=\gamma_{\mu},\,i\gamma_{\mu}\gamma_{5},\\
1 ,\:\: &\Gamma=\sigma_{\mu\nu}.
\end{array}
\right.
\end{eqnarray}

Combining these results as well as that for $\varphi^{\text{DRED}}(t)$ given in Eq.~(\ref{Fermion_ren_t}), we obtain the matching coefficients for the quark bilinear operators~$\overline{\psi}_1 \Gamma \psi_2$, 
\begin{eqnarray}
Z^{\text{GF}\to\overline{\text{MS}};\text{DRED}}_{\Gamma}(t)
&=&\frac{Z^{\overline{\text{MS}};{\text{DRED}}}_{\Gamma}}{(1+I^{\text{GF}}_{\Gamma}(t))}\frac{Z^{\overline{\text{MS}};{\text{DRED}}}_{\psi}}{\varphi^{\text{DRED}}(t)}
\nonumber \\
&=&
\left\{
\begin{array}{ll}
1+\frac{\varg^2(\mu)}{(4\pi)^2}C_F\left\{3\gamma_E+3\log(2t\Tilde\mu^2)+3+\log(432)\right\} ,\:\: &\Gamma=\1,\,\gamma_5,\\
1+\frac{\varg^2(\mu)}{(4\pi)^2}C_F\left\{\log(432)\right\} ,\:\: &\Gamma=\gamma_{\mu},\,i\gamma_{\mu}\gamma_{5},\\
1+\frac{\varg^2(\mu)}{(4\pi)^2}C_F\left\{-\gamma_E-\log(2t\Tilde\mu^2)-1+\log(432)\right\} ,\:\: &\Gamma=\sigma_{\mu\nu}.
\end{array}
\right.
\nonumber \\
\label{eq:matching_coefficient_bilinear}
\end{eqnarray}
We confirm that the $1/\epsilon$ divergences in $I^{\text{GF}}_{\Gamma}(t)$ etc.\ cancel out with each other in the combination of the matching coefficient~$Z^{\text{GF}\to\overline{\text{MS}};\text{DRED}}_{\Gamma}(t)$.

\section{Summary and outlook}
\label{sec:conclusions}

In this paper we computed the matching coefficient~$Z^{\text{GF}\to\overline{\text{MS}};{\text{DRED}}}_{O_{\pm}}(t)$ for four quark operators~$O_\pm$ defined by~Eq.~(\ref{def-op}), 
and $Z^{\text{GF}\to\overline{\text{MS}};\text{DRED}}_{\Gamma}(t)$ for quark bilinear operators~$\overline{\psi}_1 \Gamma \psi_2$ defined by~Eq.~(\ref{eq:bilinear}), adopting the dimensional reduction scheme.
Our results for the one-loop matching coefficients are given by~Eqs.~(\ref{eq:matching_coefficient_Opm}) and~(\ref{eq:matching_coefficient_bilinear}), respectively.    
Combining these results, we also obtain the matching coefficient for the kaon bag parameter $B_K$ defined by~Eq.~(\ref{eq:BK}),
\begin{eqnarray}
Z^{\text{GF}\to\overline{\text{MS}};\text{DRED}}_{B_K}(t)
&=&\frac{Z^{\text{GF}\to\overline{\text{MS}};\text{DRED}}_{O_{+}}(t)}{\left(Z^{\text{GF}\to\overline{\text{MS}};\text{DRED}}_{\gamma_\mu\gamma_5}(t)\right)^2}\nonumber\\
&=&1+\frac{\varg^2(\mu)}{(4\pi)^2}\,\frac{-3N+3}{N}\,\left(\log{(2t{\Tilde\mu}^2)}+\gamma_E+1\right),\label{matching_factor_BK}
\end{eqnarray}
where $N=3$ for QCD.
We are planning to perform simulations to study the kaon bag parameter by the SF\textit{t}X method, adopting nonperturbatively $\mathcal{O}(a)$-improved dynamical Wilson quarks~\cite{taniguchiLat19}.

These matching coefficients are important in evaluating the $\overline{\text{MS}}$ renormalized operators in the dimensional reduction scheme at the renormalization scale~$\Tilde\mu$, from corresponding lattice operators measured at small flow time~$t$ of the gradient flow.
A conventional choice for~$\Tilde\mu$ is 2 GeV.
On the other hand, we are free to choose the renormalization scale~$\mu$ for the matching of $\overline{\text{MS}}$ and gradient flow schemes, provided that the perturbative expansions are well converged. 
Some conventional choices for $\mu$ are $\mu_d=1/\sqrt{8t}$~\cite{GF_04} and $\mu_0 \equiv 1/\sqrt{2e^{\gamma_E}t}$~\cite{EMT-2loop1}, which are natural scales for flowed operators because the smearing range is $\sim \sqrt{8t}$ by the gradient flow.
In practice, however, because the perturbative expansions are truncated, the quality of the results may be affected by the choice of~$\mu$. 
Recently, we found that an optimal choice of $\mu$ can improve the reliability and applicability of the SF\textit{t}X method~\cite{EMT-2loop2}. 
Such improvement may be important in evaluating complicated operators, such as $O_\pm$ for the kaon bag parameter.

\acknowledgments
We thank the members of the WHOT-QCD Collaboration for valuable discussions.
This work is in part supported by JSPS KAKENHI Grants No. JP16H03982, No. JP18K03607, No. JP19K03819, and No. JP20H01903.


\begin{thebibliography}{99}

\bibitem{history} 
P.~Dimopoulos, arXiv:1101.3069;
S.~Aoki \textit{et al.}, Eur. Phys. J. C \textbf{80}, 113 (2020).
\bibitem{BK_01}M.B.~Gavela, L.~Maiani, S.~Petrarca, F.~Rapuano \textit{et al.}, Nucl.\ Phys.\ B\textbf{306}, 677 (1988).
\bibitem{BK_02}C.~Bernard and A.~Soni, Nucl.\ Phys.\ B, Proc. Suppl. \textbf{17} , 495 (1990); \textbf{42}, 391 (1995).
\bibitem{BK_03}R.~Gupta, D.~Daniel, G.W.~Kilcup, A.~Patel, and S.R.~Sharpe, Phys.\ Rev.\ D \textbf{47}, 5113 (1993).
\bibitem{BK_04}A.~Donini, G.~Martinelli, C.T.~Sachrajda, M.~Talevi, and A.~Vladikas, Phys.\ Lett.\ B \textbf{360}, 83 (1995).
\bibitem{BK_05}M.~Crisafulli, A.~Donini, V.~Lubicz, G.~Martinelli, F.~Rapuano, M.~Talevi, C.~Ungarelli, and A.~Vladikas, Phys.\  Lett.\ B \textbf{369}, 325 (1996).
\bibitem{BK_06}A.~Donini, V.~Gim\'enez, G.~Martinelli, G.C.~Rossi, M.~Talevi, M.~Testa, and A.~Vladikas, Nucl.\ Phys.\ B, Proc.\ Suppl. \textbf{53}, 883 (1997).
\bibitem{BK_07}L.~Conti, A.~Donini, V.~Gimenez, G.~Martinelli, M.~Talevi, and A.~Vladikas, Nucl. Phys. B, Proc. Suppl. \textbf{63}, 880 (1998); 
Phys.\ Lett.\ B \textbf{421}, 273 (1998).
\bibitem{BK_08}R.~Gupta, T.~Bhattacharya and S.~Sharpe, Phys. Rev. D \textbf{55}, 4036 (1997).
\bibitem{BK_09}R.~Gupta, Nucl.Phys. B (Proc.Suppl.) \textbf{63}, 278 (1998).
\bibitem{BK_10}S.~Aoki, M.~Fukugita, S.~Hashimoto, N.~Ishizuka, Y.~Iwasaki, K.~Kanaya, Y.~Kuramashi, M.~Okawa, A.~Ukawa, and T.~Yoshi\'e (JLQCD Collaboration), Nucl. Phys. B, Proc.Suppl. \textbf{60}A, 67 (1998); Phys. Rev. D \textbf{60}, 034511 (1999).

\bibitem{GF_00}R.~Narayanan and H.~Neuberger,
  J.\ High Energy Phys.\ 03 (2006) 064.
\bibitem{GF_03}M.~L\"uscher, Commun. Math. Phys. \textbf{293}, 899 (2010).
\bibitem{GF_04}M.~L\"uscher, J.\ High Energy Phys.\ 08 (2010) 071; 03 (2014) 092(E).
\bibitem{GF_01}M.~L\"uscher and P.~Weisz, J.\ High Energy Phys.\ 02 (2011) 051.
\bibitem{GF_02}M.~L\"uscher, J.\ High Energy Phys.\ 04 (2013) 123.
\bibitem{GF_05}M.~L\"uscher, Proc. Sci., LATTICE2013 (2014) 016 [arXiv:1308.5598].

\bibitem{GF_ap01}H.~Suzuki, Prog.\ Theor.\ Exp.\ Phys.\  {\bf 2013}, 083B03 (2013); {\bf 2015}, 079202(E) (2015)].
\bibitem{GF_ap03}H.~Makino and H.~Suzuki, Prog.\ Theor.\ Exp.\ Phys.\  {\bf 2014}, 063B02 (2014); {\bf 2015}, 079202(E) (2015).

\bibitem{GF_10} A.~Ramos,
 Proc.\ Sci., LATTICE2014 (2015) 017 [arXiv:1506.00118].
\bibitem{GF_11}H.~Suzuki,
 Proc.\ Sci., LATTICE2016 (2017) 002 [arXiv:1612.00210].

\bibitem{GF_12}J.~Artz, R.V.~Harlander, F.~Lange, T.~Neumann, and M.~Prausa, J. High Energy Phys. {\bf 06} (2019) 121.
 
\bibitem{Tec_01}T.~Endo, K.~Hieda, D.~Miura and H.~Suzuki, Prog.\ Theor.\ Exp.\ Phys.\  {\bf 2015}, 053B03 (2015).

\bibitem{Tec_02}H.~Suzuki, Prog.\ Theor.\ Exp.\ Phys.\  {\bf 2015}, 103B03 (2015).

\bibitem{Tec_04}K.~Hieda and H.~Suzuki, Mod.\ Phys.\ Lett.\ A \textbf{31}, 1650214 (2016).

\bibitem{Asakawa:2013laa} 
  M.~Asakawa, T.~Hatsuda, E.~Itou, M.~Kitazawa, and H.~Suzuki (FlowQCD Collaboration),
  Phys.\ Rev.\ D {\bf 90}, 011501 (2014),
  ;{\bf 92}, 059902(E) (2015).

\bibitem{FlowQCD1}
  M.~Kitazawa, T.~Iritani, M.~Asakawa, T.~Hatsuda, and H.~Suzuki, 
  Phys.\ Rev.\ D {\bf 94}, 114512 (2016).

\bibitem{Iritani2019}
T.~Iritani, M.~Kitazawa, H.~Suzuki, and  H.~Takaura, 
Prog.\ Theor.\ Exp.\ Phys.\ {\bf 2019}, 023B02 (2019).

\bibitem{GF_ap08}Y.~Taniguchi, S.~Ejiri, R.~Iwami, K.~Kanaya, M.~Kitazawa, H.~Suzuki, T.~Umeda and N.~Wakabayashi, Phys.\ Rev.\ D \textbf{96}, 014509 (2017).

\bibitem{GF_ap09}Y.~Taniguchi, K.~Kanaya, H.~Suzuki, and T.~Umeda, Phys.\ Rev.\ D \textbf{95}, 054502 (2017).

\bibitem{Lat2017-kanaya}
K.~Kanaya , S.~Ejiri, R.~Iwami, M.~Kitazawa, H.~Suzuki, Y.~Taniguchi, and T.~Umeda, 
EPJ Web Conf.\ {\bf 175}, 07023 (2018).

\bibitem{EMT-2loop1}R.V.~Harlander, Y.~Kluth, and F.~Lange, Eur. Phys. J. C 78, 944 (2018).

\bibitem{EMT-2loop2}
Y.~Taniguchi, S.~Ejiri, K.~Kanaya, M.~Kitazawa, H.~Suzuki and T.~Umeda, 
Phys. Rev. D \textbf{102}, 014510 (2020).

\bibitem{BK_11}S.~Aoki, M.~Fukugita, S.~Hashimoto, N.~Ishizuka, Y.~Iwasaki, K.~Kanaya, Y.~Kuramashi, M.~Okawa, A.~Ukawa, and T.~Yoshi\'e (JLQCD Collaboration), Nucl.\ Phys.\ B, Proc.Suppl. \textbf{60A}, 67 (1998) ; Phys.\ Rev.\ D \textbf{60}, 034511 (1999).

\bibitem{sOPE}C.~Monahan and K.~Orginos, Phys. Rev. D \textbf{91}, 074513 (2015).

\bibitem{GF_ap10}M.~Rizik, C.~Monahan, and A.~Shindler, Proc.\ Sci., LATTICE2018 (2019) 215 [arXiv:1810.05637 [hep-lat]].

\bibitem{GF_ap11}J.G.~Reyes, J.~Dragos, J.~Kim, A.~Shindler, and T.~Luu, Proc.\ Sci., LATTICE2018 (2019) 219 [arXiv:1811.11798]
.
\bibitem{GF_ap12}J.~Kim, J.~Dragos, A.~Shindler, T.~Luu, and J.~de~Vries, Proc.\ Sci., LATTICE2018 (2019) 260 [arXiv:1810.10301].

\bibitem{GF_ap13}M.D.~Rizik, C.J.~Monahan and A.~Shindler, [arXiv:2005.04199].

\bibitem{Twisted_mass01}M.~Constantinou, P.~Dimopoulos, R.~Frezzotti, K.~Jansen, V.~Gimenez, V.~Lubicz, F.~Mescia, H.~Panagopoulos, M.~Papinutto, G.C.~Rossi, S.~Simula, A.~Skouroupathis, F.~Stylianou, and A.~Vladikas, Phys.\ Rev.\ D \textbf{83}, 014505 (2011).

\bibitem{Twisted_mass02}N. Carrasco, P. Dimopoulos, R. Frezzotti, V. Lubicz, G.C. Rossi, S. Simula, and C. Tarantino, Phys.\ Rev.\ D \textbf{92}, 034516 (2015).


\bibitem{DRED} W.~Siegel, Phys.\ Lett.\ B \textbf{84}, 193 (1979).

\bibitem{MS-bar} W.A.~Bardeen, A.J.~Buras, D.~W.~Duke and T.~Muta, Phys.\ Rev.\ D \textbf{18}, 3998 (1978).

\bibitem{Evanescent01}A.J.~Buras and P.~H.~Weisz, Nucl.\ Phys.\ \textbf{B333}, 66 (1990).
\bibitem{Evanescent02}S.~Herrlich and U~ Nierste, Nucl.\ Phys.\ \textbf{B455}, 39 (1995).

\bibitem{Tec_05}G. Martinelli, Phys.\ Lett.\ B \textbf{141}, 395 (1984).

\bibitem{Tec_07}Y. Taniguchi, J.\ High Energy Phys.\ {\bf 1204} (2012) 143.

\bibitem{taniguchiLat19}Y.~Taniguchi, \textit{Proceedings of the 37th International Symposium on Lattice Field Theory (Lattice 2019), Wuhan, China (unpublished).}

\end{thebibliography}
\end{document}